\documentclass[aps,prb,twocolumn,superscriptaddress,floatfix,longbibliography]{revtex4-1}

\usepackage{physics}
\usepackage{graphicx}
\usepackage{booktabs}
\usepackage{bm, bbm, amssymb, siunitx, gensymb}

\usepackage{xcolor}
\definecolor{mblue}{RGB}{0, 118, 186}

\definecolor{mgreen}{RGB}{29, 177, 0}

\definecolor{mred}{RGB}{209, 28, 36}

\usepackage{hyperref}
\hypersetup{
    colorlinks=true,
    linkcolor=blue,     
    urlcolor=blue,
    citecolor=blue
}
\newif\ifptitle
\newif\ifpnumber
\newcounter{para}
\newcommand\ptitle[1]{\par\refstepcounter{para}
{\ifpnumber{\noindent\textcolor{darkgray}{\textbf{\thepara}}\indent}\fi}
{\ifptitle{\textbf{[{#1}]}}\fi}}
\pnumbertrue  

\begin{document}

\title{Acoustic twisted bilayer graphene}
    
\author{S.~Minhal Gardezi}
\affiliation{School of Engineering and Applied Sciences, Harvard University, Cambridge, MA, 02138, USA}
\author{Harris Pirie}
\affiliation{Department of Physics, Harvard University, Cambridge, MA, 02138, USA}
\author{William Dorrell}
\affiliation{Department of Physics, Harvard University, Cambridge, MA, 02138, USA}
\author{Jennifer E. Hoffman}
\affiliation{School of Engineering and Applied Sciences, Harvard University, Cambridge, MA, 02138, USA}

\affiliation{Department of Physics, Harvard University, Cambridge, MA, 02138, USA}

\date{\today}

\begin{abstract}
Twisted van der Waals (vdW) heterostructures have recently emerged as an attractive platform to study tunable correlated electron systems. However, the quantum mechanical nature of vdW heterostructures makes their theoretical and experimental exploration laborious and expensive. Here we present a simple platform to mimic the behavior of twisted vdW heterostructures using acoustic metamaterials comprising of interconnected air cavities in a steel plate. Our classical analog of twisted bilayer graphene shows much of the same behavior as its quantum counterpart, including mode localization at a magic angle of $\sim1.1\degree$. By tuning the thickness of the interlayer membrane, we reach a regime of strong interactions more than three times higher than the feasible range of twisted bilayer graphene under pressure. In this regime, we find the magic angle as high as $6.01\degree$, corresponding to a far denser array of localized modes in real space and further increasing their interaction strength. Our results broaden the capabilities for cross-talk between quantum mechanics and acoustics, as vdW metamaterials can be used both as simplified models for exploring quantum systems and as a means for translating interesting quantum effects into acoustics. 
\end{abstract}

\maketitle

\newpage

\ptitle{Twistronics vastly expands the vdW phase space}
Van der Waals (vdW) heterostructures host a rich range of emergent properties that can be customized by varying the stacking configuration of sheets of two-dimensional materials, such as graphene, other xenes, or transition-metal dichalcogenides \cite{Geim-Grigorieva2013Nature,Ajayan-Banerjee2016PhysToday,Novoselov-CastroNeto2016Science}. As the library of two-dimensional materials expands, vdW heterostructures continue to present new ways to simplify and broaden existing technology \cite{LiangAdv.Mater.2019}. Recently, these possibilities have been expanded enormously by considering the addition of a small twist angle between adjacent layers in a vdW heterostructure, leading to the growing field of twistronics \cite{Carr-Kaxiras2017PRB}. The twist angle induces a moir\'{e} pattern that acts as a highly tunable potential for electrons moving in the layers. In combination with the intrinsic interlayer coupling, this moir\'{e} potential can manipulate the velocity of itinerant charge carriers, leading to enhanced electron correlations when their kinetic energy is reduced below their Coulomb interaction. Even traditional non-interacting materials can reach this regime, as evidenced by the discovery of a correlated insulating state in twisted bilayer graphene \cite{Cao-JarilloHerrero2018Nature1}. Already, vdW heterostructures with moir\'{e} superlattices have led to the development of new platforms for Wigner crystals \cite{ReganNature2020}, interlayer excitons \cite{JinNature2019, SeylerNature2019, TranNature2019}, and unconventional superconductivity \cite{Cao-JarilloHerrero2018Nature2, Yankowitz-Dean2019Science, ChenNature2019}. But the search for such novel twistronic phases is still in its infancy and there are countless vdW stacking and twisting arrangements that remain unexplored. Moreover, much of the correlated physics occurs at small twist angles, where traditional density functional theory calculations are challenging due to the large size of the moir\'{e} supercell. Consequently, it remains pressing to develop a platform to rapidly prototype and explore new twistronic materials to accelerate their technological advancement.

\begin{figure*}
  \includegraphics[width=0.8\textwidth]{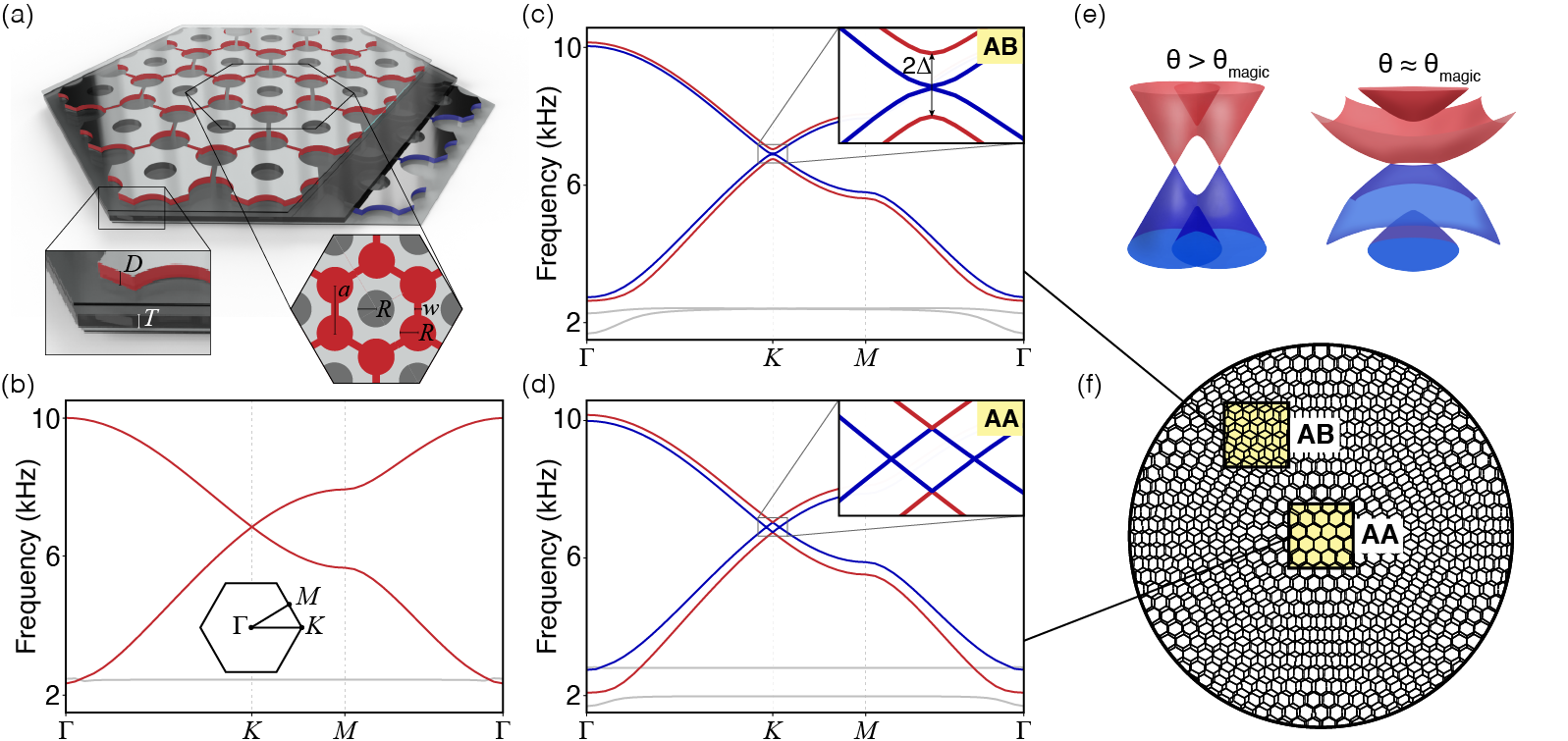}
  \caption{Acoustic vdW metamaterials. 
  (a) In this acoustic metamaterial, sound waves propagate through connected air cavities in solid steel to mimic the way that electrons hop between carbon atoms in bilayer graphene. Our metamaterial graphene has cavity spacing $a=10$ mm, cavity radius $R=3.5$ mm, channel width $w=0.875$ mm, and steel thickness $D=1$ mm. The presence of smaller, unconnected cavities in the center of each unit cell improve the interlayer coupling of our bilayer metamaterial, but do not act as additional lattice sites. 
  (b) The $C_6$ symmetry of the lattice protects a Dirac-like crossing at the $K$ point of the monolayer acoustic band structure. 
  (c-d) Two sheets of acoustic graphene coupled by an interlayer polyethylene membrane of thickness $T = 2.21$ mm accurately recreates the AB and AA configurations of bilayer graphene. 
  (e) As the two sheets are twisted relative to one another, the two Dirac cones contributed by each layer hybridize, ultimately producing a flat band at small twist angles. 
  (e) The twisted bilayer heterostructure has its own macroscopic periodicity with distinct AB and AA stacking regions.}
  \label{f1}
\end{figure*}

\ptitle{Acoustic metamaterials as quantum mimics}
The development of acoustic metamaterials over the last few years has unlocked a compelling platform to guide the design of new quantum materials \cite{GeNatl.Sci.Rev.2018}. Whereas quantum materials can be difficult to predict and fabricate, acoustic metamaterials have straightforward governing equations, easily adjustable properties, and fast build times, making them attractive models to rapidly explore their quantum counterparts. A well-designed acoustic metamaterial can redirect sound waves to track the motion of electrons moving through crystalline solids. Nowadays, cutting-edge ideas in several areas of theoretical condensed matter are demonstrated using phononic analogs, such as chiral Landau levels \cite{PeriNat.Phys.2019}, higher-order topology \cite{Serra-GarciaNature2018, NiNatureMater2019}, and fragile topology \cite{PeriScience2020}. The utility of phononic metamaterials also spans a large range of vdW heterostructures: the Dirac-like electronic bands in graphene have been mimicked using longitudinal acoustics \cite{MeiPhys.Rev.B2012, TorrentPhys.Rev.Lett.2012, Lu-Liu2014PRB}, surface acoustics waves \cite{YuNatureMater2016}, and mechanics \cite{TorrentPhysRevB2013, KariyadoSciRep2016}. Further, it was recently discovered that placing a thin membrane between metamaterial layers can reproduce the coupling effects of vdW forces, yielding acoustic analogs of bilayer and trilayer graphene \cite{LuPhysRevLett2018, Dorrell-Hoffman2020PRB}. The inclusion of a twist angle between metamaterial layers has the potential to expand their utility even further. In addition to electronic systems, moir\'{e} engineering has recently been demonstrated in systems containing vibrating plates \cite{RosendoArXiv200811706Cond-Mat2020}, spoof surface-acoustic waves \cite{DengArXiv201005940Cond-Mat2020}, and optical lattices \cite{WangNature2020}. However, without simultaneous control of intralayer hopping, interlayer coupling, and twist angle, rapidly prototyping next-generation twistronic devices using acoustic metamaterials remains an elusive goal. 

\ptitle{Here we show}
Here we introduce a simple acoustic metamaterial that accurately recreates the band structure of twisted bilayer graphene, including enhanced mode localization at a magic angle of 1.12$\degree$.  Our bottom-up approach begins with a straightforward monolayer acoustic metamaterial that implements a tight-binding model describing the low-energy band structure of graphene. By separating two of these monolayers with a thin polyethylene membrane, we recreate both stacking configurations of bilayer graphene in our finite-element simulations using {\sc comsol multiphysics}. Our acoustic analog of bilayer graphene hosts flat bands at the same magic angle of $\sim1.1\degree$ as its quantum counterpart \cite{Bistritzer-MacDonald2011PNAS, Cao-JarilloHerrero2018Nature1}. The advantage of our metamaterial is the ease with which it reaches coupling conditions beyond the feasible capabilities of conventional bilayer graphene. By tuning the thickness of the interlayer membrane, we simulate new metamaterials that host flat bands at several magic angles between 1.12$\degree$ and 6.01$\degree$. Our results demonstrate the potential for vdW metamaterials to simulate and explore the ever-growing number of quantum twistronic materials.

\ptitle{Using air cavities as atoms}
We begin by introducing a general framework to acoustically mimic electronic materials that are well described by a tight-binding model, in which the electron wavefunctions are fairly localized to the atomic sites. In our metamaterial, each atomic site is represented by a cylindrical air cavity in a steel plate (see Fig.~\hyperref[f1]{1(a)}). The radius of the air cavity determines the frequencies of its ladder of acoustic standing modes. In a lattice of these cavities, the degenerate standing modes form narrow bands, separated from each other by a large frequency gap. We primarily focus on the lowest, singly degenerate $s$ band. Just as electrons hop from atom to atom in an electronic tight-binding model, sound waves propagate from cavity to cavity in our acoustic metamaterial through a network of tunable thin air channels. This coupling is always positive for $s$ cavity modes, but either sign can be realized by starting with higher-order cavity modes \cite{MatlackNatureMater2018}. Because sound travels much more easily through air than through steel, these channels are the dominant means of acoustic transmission through our metamaterial. They allow nearest- and next-nearest-neighbor coupling to be controlled independently by varying the width or length of separate air channels, providing a platform to implement a broad class of tight-binding models.

\ptitle{Monolayer and bilayer graphene}
To recreate bilayer graphene in an acoustic metamaterial, we started from a honeycomb lattice of air cavities, with radius of 3.5 mm and a separation of 10 mm, in a 1-mm-thick steel sheet, encapsulated by 1-mm-thick polyethylene boundaries. Each cavity is coupled to its three nearest neighbors using 0.875-mm-wide channels, giving a $s$-mode bandwidth of $2t=7.8$ kHz (see Fig.~\hyperref[f1]{1(b)}). This $s$ manifold is well isolated from other higher-order modes in the lattice, which appear above 25 kHz. The $C_6$ symmetry of our metamaterial ensures a linear crossing at the $K$ point, similar to the Dirac cone in graphene \cite{Kogan-Nazarov2012PRB}.  The frequency of this Dirac-like crossing and other key aspects of the band structure are easily tunable by varying the dimensions of the cavities and channels. Building on previous work, we coupled two layers of acoustic graphene together using a thin interlayer membrane \cite{Dorrell-Hoffman2020PRB}. By adjusting only the stacking configuration, the same acoustic metamaterial mimics both the parabolic touching around the $K$ point seen in AB-stacked bilayer graphene, and the offset Dirac bands seen in AA-stacked bilayer graphene \cite{deAndres-Verges2008PRB, McCann-Koshino2013RPP}, as shown in Fig.~\hyperref[f1]{1(c-d)}. The vertical span between the $K$-point eigenmodes gives the interlayer coupling strength $2\Delta$, which is adjustable by changing the membrane?s thickness \cite{Dorrell-Hoffman2020PRB}. We found that a 2.21-mm-thick polyethylene membrane (density 950 kg/m$^3$ and speed of sound 2460 m/s) accurately matched the dimensionless coupling ratio $\Delta/t \approx 5\%$ in bilayer graphene. 

\begin{figure}[t]
  \includegraphics[width=0.48\textwidth]{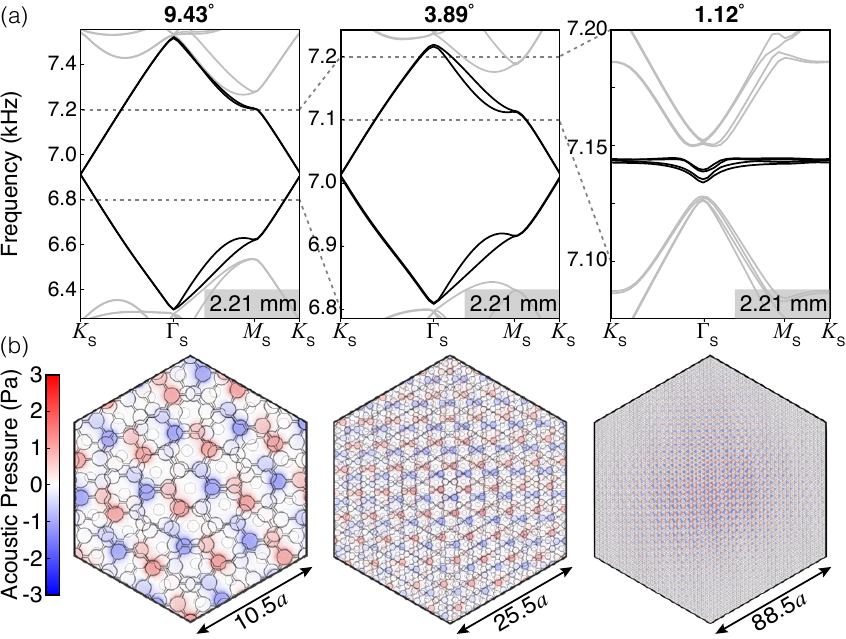}
  \caption{Acoustic flat band at small angles. (a) As the twist angle between two layers of acoustic graphene decreases, its supercell shrinks to fold the Dirac-like bands to a narrower frequency interval. Eventually at a magic angle of 1.12$\degree$, the uppermost band becomes almost completely flat. (b) In real space, the Dirac-like modes at the $K_\mathrm{s}$ point are itinerant and persist across the entire supercell for large angles. But at the magic angle, they become localized on the AA-stacked region in the center of the supercell.  }
  \label{f2}
\end{figure}

\ptitle{Acoustic magic angles}
Introducing a twist angle between two graphene layers creates a moir\'{e} pattern that grows in size as the angle decreases. A large moir\'{e} lattice corresponds to a small supercell Brillouin zone, in which the Dirac cone from one layer is folded onto the $K_\mathrm{s}$ point, while that of the other layer is folded to the $K^\prime_\mathrm{s}$ point.  As twist angle decreases, these two Dirac cones are pushed together and hybridize due to the interlayer coupling \cite{Shallcross-Pankratov2010PRB, Moon-Koshino2013PRB}, as shown in Fig.~\hyperref[f1]{1(e)}. Eventually, they form a flat band with a vanishing Fermi velocity at a so-called magic angle \cite{SuarezMorell-Barticevic2010PRB,Bistritzer-MacDonald2011PNAS,Cao-JarilloHerrero2018Nature1}. We searched for the same band-flattening mechanism by introducing a commensurate-angle twist to our acoustic bilayer graphene metamaterial. Beginning from untwisted AA stacking, we discovered hybridized Dirac bands with widths that decrease at small twist angles (Fig.~\hyperref[f1]{2(a)}), matching the flattening trend found in twisted bilayer graphene \cite{Moon-Koshino2013PRB}. Strikingly, our metamaterial mimics its quantum counterpart even down to the magic angle, producing acoustic flat bands at $1.12\degree$. These flat bands correspond to real-space pressure modes that become primarily localized on the AA region, see Fig.~\hyperref[f2]{2(b)}. This AA localization agrees with calculations of the local density of states in magic-angle twisted bilayer graphene \cite{Kim-Tutuc2017PNAS, Cao-JarilloHerrero2018Nature1}. In our acoustic system, it represents sounds waves that propagate with a low group velocity of 0.03 m/s compared to 30 m/s in the untwisted bilayer. 


\begin{figure}[t]
  \includegraphics[width=0.48\textwidth]{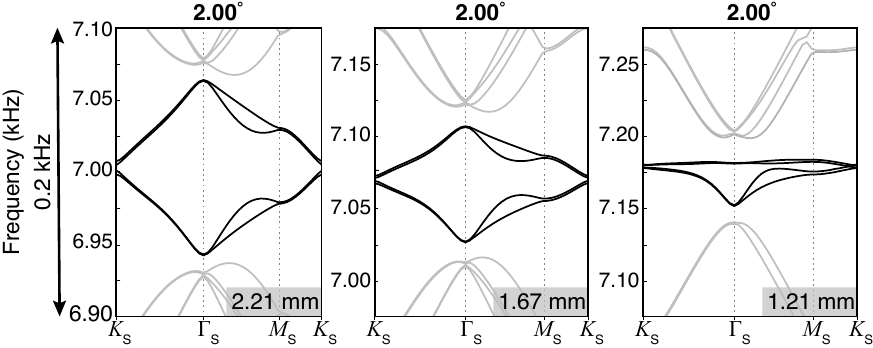}
  \caption{Interlayer coupling tunes the magic angle. 
  Reducing the thickness of the polyethylene membrane enhances the coupling between layers of our acoustic metamaterial and consequently increase its magic angle. For example, to realize flat bands at 2$\degree$, we need to reduce the membrane thickness to 1.21 mm. }
  \label{f3}
\end{figure}

\ptitle{Interlayer coupling tunes the magic angle}
The magic angle of twisted bilayer graphene can be tuned by applying vertical pressure \cite{Carr-Kaxiras2018PRB}. Intuitively, greater pressure pushes the graphene layers together and increases the interlayer coupling strength. Consequently, the Dirac bands begin to flatten at higher angles than at ambient pressure. However, a substantial vertical pressure of a few GPa is required to move the magic angle from 1.1$\degree$ to 1.27$\degree$, corresponding to only a 20\% increase in the interlayer coupling strength \cite{Yankowitz-Dean2019Science}. In our metamaterial, no such physical restrictions apply: the interlayer coupling can be tuned continuously over two orders of magnitude simply by changing the thickness of the coupling membrane \cite{Dorrell-Hoffman2020PRB}. To demonstrate this flexibility, we reproduced the band flattening at a fixed angle of 2$\degree$ simply by incrementally reducing the thickness of the interlayer (Fig.~\hyperref[f3]{3}). The bands flattened completely for a 1.21-mm-thick interlayer, making 2$\degree$ a magic angle for such a metamaterial. In other words, the flat band condition can be approached from two directions: either reduce the twist angle at fixed interlayer thickness, or reduce the thickness at fixed angle. 

\begin{figure}[t]
  \includegraphics[width=0.48\textwidth]{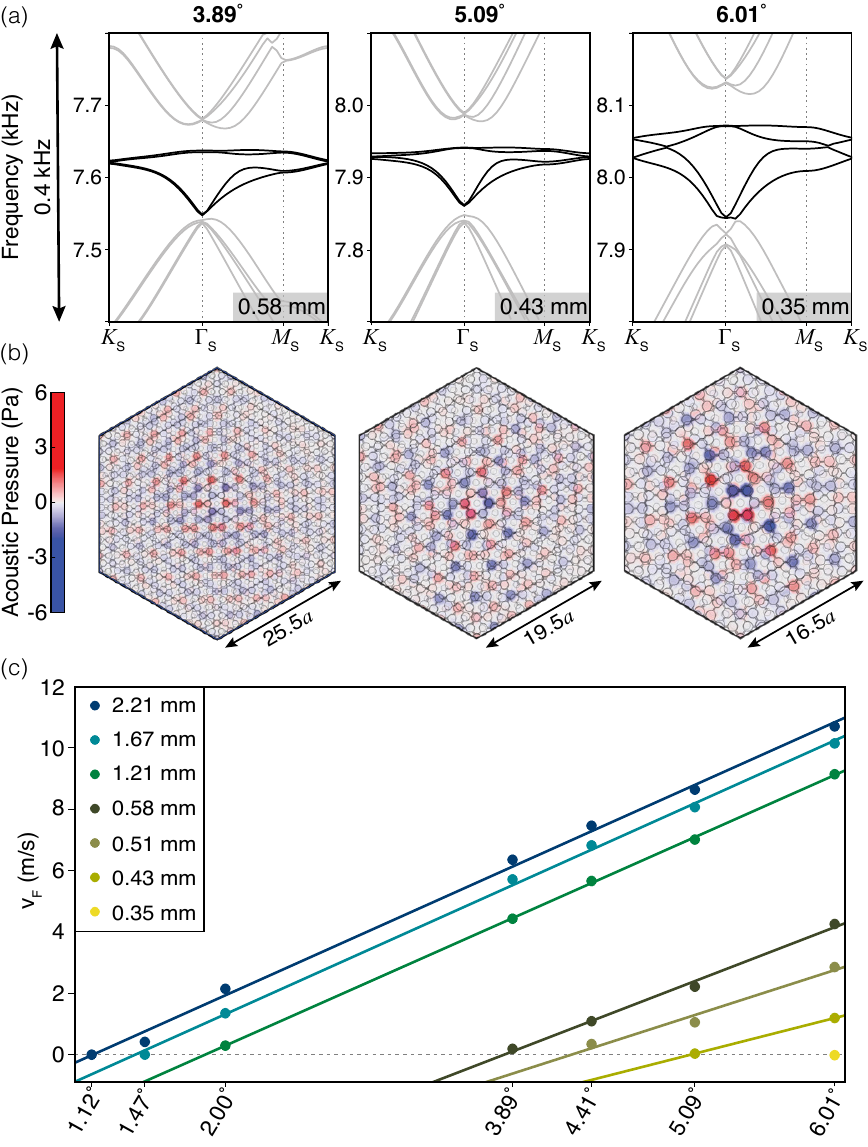}
  \caption{Reproducing flat bands at high magic angles. 
  (a) By reducing the thickness of the interlayer membrane to 0.58 mm, 0.43 mm, and 0.35 mm, we discover flat bands at 3.89$\degree$, 5.09$\degree$, and 6.01$\degree$ twist angles. 
  (b) In each case, the $K_\mathrm{s}$-point eigenmodes appear predominantly around the AA-stacked region in the center of the supercell. As the supercell shrinks, these localized modes form a dense array, increasing the possibility of interactions with similar localized modes in neighboring supercells.
  (c) By correctly choosing the interlayer membrane's thickness, any angle can become a magic angle with a vanishing $K_\mathrm{s}$-point velocity.}
  \label{f4}
\end{figure}

\ptitle{Coupling regimes beyond graphene}
Our twisted bilayer metamaterial provides a simple platform to explore twistronics in extreme coupling regimes, well beyond the experimental capability of its electronic counterpart. By reducing the interlayer membrane thickness, we searched for flat bands at high commensurate angles of 3.89$\degree$, 5.09$\degree$, and 6.01$\degree$. In each case, we discovered a band structure that hosts flat bands similar to those in twisted bilayer graphene (Fig.~\hyperref[f4]{4(a)}). These flat bands all correspond to collective pressure modes that are localized on the AA-stacked central regions (Fig.~\hyperref[f4]{4(b)}). But these localized modes are over five times closer to each other at $6.01\degree$ than they are at $1.12\degree$. Generally speaking, the modes interact more strongly as they get closer together, potentially allowing interactions to dominate over the reduced kinetic energy at a high magic angle. Consequently, a high-magic-angle metamaterial could be susceptible to non-linear effects if tuned correctly, akin to a phonon-phonon interaction. In principle, any twist angle can become a magic angle that hosts a dense array of such localized modes, by choosing the correct interlayer thickness (Fig.~\hyperref[f4]{4(c)}). 

\ptitle{Conclusion}
Our twisted bilayer metamaterial represents a successful translation of twistronics into the field of acoustics. By introducing a twist angle to vdW metamaterials, we discovered flat bands that mimic the behavior of twisted bilayer graphene at 1.1$\degree$ and slow transmitted sound by a factor of a thousand (40$\times$ slower than a leisurely walk). Although we focused on analogs of graphene, the band structures of many other two-dimensional materials, such as bismuthene or hexagonal boron nitride, can be captured in a simple tight-binding model and described by our cavity metamaterial.  In addition, the dense lattice of localized modes at higher magic angles may lead to new paths for acoustic emitters and ultrasound imagers. Finally, our design of magic-angle twisted vdW metamaterials directly translates to photonic systems under a simple mapping of variables \cite{MeiPhys.Rev.B2012}.\\

\section*{Acknowledgements}

We thank Alex Kruchkov, Stephen Carr, Clayton DeVault, Haoning Tang, Fan Du, Nathan Drucker, Ben November, and Walker Gillett for insightful discussions. This work was supported by the Science and Technology Center for Integrated Quantum Materials under the National Science Foundation Grant No. DMR-1231319.

\vspace{5cm}
\bibliography{refs.bib}

\end{document}